\documentclass[
reprint,
prl,
superscriptaddress,
showpacs,
amsmath,
amssymb,
]{revtex4-2}

\usepackage{lineno}
\usepackage{mathtools}
\usepackage{color,soul}
\usepackage{graphicx}
\usepackage{dcolumn}
\usepackage{bm}
\usepackage{feynmp-auto}
\usepackage{amssymb,amsfonts,amsmath}
\usepackage{threeparttable}
\usepackage{textcomp}
\usepackage{float}
\usepackage{wrapfig}
\usepackage{verbatim}
\usepackage{accents}
\usepackage{fancyhdr}
\usepackage{euscript}
\usepackage{blkarray}
\usepackage{multirow}
\usepackage{tikz}
\usepackage{subfigure}
\usepackage{enumitem}
\usepackage{physics}
\usepackage[T1]{fontenc}
\usepackage{mathptmx}
\usetikzlibrary[topaths]
\usepackage[colorlinks,
            linkcolor=red,
            anchorcolor=blue,
            citecolor=blue
            ]{hyperref}

\begin{document}
\title{Negative Thermal Hall Conductance in Two-Dimer Shastry-Sutherland Model with $\pi$-flux Dirac Triplon}

\author{Hao Sun}
\email{sunhao@ntu.edu.sg}
\affiliation{School of Electrical and Electronic Engineering, Nanyang Technological University, 50 Nanyang Avenue, Singapore 639798, Singapore}

\author{Pinaki Sengupta}
\email{psengupta@ntu.edu.sg}
\affiliation{School of Physical and Mathematical Sciences, Nanyang Technological University, 21 Nanyang Link, Singapore 637371, Singapore}

\author{Donguk Nam}
\email{dnam@ntu.edu.sg}
\affiliation{School of Electrical and Electronic Engineering, Nanyang Technological University, 50 Nanyang Avenue, Singapore 639798, Singapore}

\author{Bo Yang}
\email{yang.bo@ntu.edu.sg}
\affiliation{School of Physical and Mathematical Sciences, Nanyang Technological University, 21 Nanyang Link, Singapore 637371, Singapore}
\affiliation{Institute of High Performance Computing, A*STAR, Singapore, 138632.}
 
\date{\today}

\begin{abstract}
We introduce an effective 2-dimer tight-binding model for the family of Shastry-Sutherland models with geometrically tunable triplon excitations. The Rashba pseudospin-orbit coupling induced by the tilted external magnetic field leads to elementary excitations having nontrivial topological properties with $\pi$-Berry flux. The interplay between the in-plane and out-of-plane magnetic field thus allows us to effectively engineer the band structure in this bosonic system. In particular, the in-plane magnetic field gives rise to Berry curvature hotspot near the bottom of the triplon band, and at the same time significantly increases the critical magnetic field for the topological triplon band. We calculate explicitly the experimental signature of the thermal Hall effect of triplons in $\rm SrCu_2(BO_3)_2$, and show a pronounced and tunabled transport signals within the accessible parameter range, particularly with a change of sign of the thermal Hall conductance. The tilted magnetic field is also useful in reducing the bandwidth of the lowest triplon band. We show it can thus be a flexible theoretical and experimental platform for the correlated bosonic topological system.
\end{abstract}

\maketitle

\begin{figure}[t]
\centering  
\includegraphics[width=0.45\textwidth]{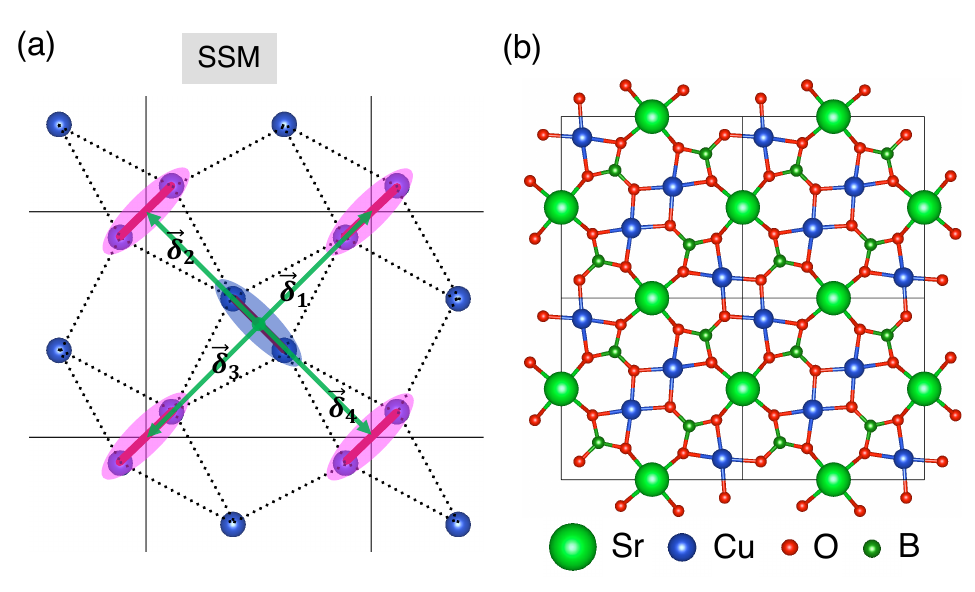}
\caption{\label{}
Lattice structure of SSM and atomic structure of $\rm SrCu_2(BO_3)_2$. (a) Lattice structure of Shastry-Sutherland model. (b) Top view of the monolayer $\rm SrCu_2(BO_3)_2$. The magnetically active $\rm Cu^{2+}$ ions form a 2D arrangement of mutually orthogonal dimers as shown in the translucent ellipses.}
\label{fig:1}
\end{figure}

{\em Introduction.}---The Shastry-Sutherland model (SSM)~\cite{Shastry1981} is a  paradigmatic Hamiltonian for studying the interplay between geometric frustration and strong interactions in quantum magnets. The canonical SSM consists of the $S=1/2$ antiferromagnetic Heisenberg model on the geometrically frustrated  Shastry-Sutherland (SS) lattice (fig.\ref{fig:1}(a)). In the strongly frustrated regime, the ground state is comprised of singlet dimers on the short bonds highlighted by the translucent bonds. Its realization in  $\rm SrCu_2(BO_3)_2$ enabled various experimental studies~\cite{Miyahara1999, Kodama2002, Sebastian2008, Jaime2012, Radtke2015, McClarty2017, McClarty2017} revealing rich many-body physics. In $\rm SrCu_2(BO_3)_2$  layers of strongly interacting $S = 1/2$ copper atoms are arranged on the SS lattice with the nearest-neighbour (NN) spin moments on $\rm Cu$ atoms forming the singlet dimers arranged in a frustrated geometry~\cite{Sebastian2008, Jaime2012,Zayed2017}. Both theoretical and experimental developments explored the possibility of exotic states including spin liquids, spin supersolids, and complex magnetic textures~\cite{Gaulin2005,Ng2006,Sengupta2007a,Sengupta2007b,Balents2010,Chen2010}. Magnetostriction and magnetocaloric measurements show a rich spectrum of magnetization plateaus and stripe-like magnetic textures in applied fields up to 100 T~\cite{Jaime2012}.  In $\rm SrCu_2(BO_3)_2$, small, but non-zero anisotropies from the Dzyaloshinskii-Moriya (DM) interactions~\cite{Cheng2007,Romhanyi2015, Kageyama1999,Knetter2000} impart a topological character to the lowest magnetic excitations. To a very good approximation, the low lying excitations in SSM are triplons that obey Bose Einstein statistics. Because of the gapped dimer-singlet ground state and triplet elementary excitations~\cite{Miyahara1999, Jaime2012, Radtke2015, McClarty2017}, exotic phases of triplons can be obtained by varying the model parameters and employing  field effects\cite{Romhanyi2015,McClarty2017,Wang2018}. The theoretically proposed gap opening has been detected by inelastic neutron scattering in a weak magnetic field, and the dimerized quantum magnet of $\rm SrCu_2(BO_3)_2$ emerges as a promising candidate to host topological bosonic phases~\cite{Romhanyi2015,McClarty2017}.

Bosonic analogs of topological phases have steadily gained interest over the past several years, and have already been proposed with photons, magnons, phonons, and skyrmionic textures~\cite{Raghu2008,Katsura2010,Zhang2010,van2013}. 
Recently, using a single dimer model, it was proposed that triplon excitations exhibit a topological phase transition in $\rm SrCu_2(BO_3)_2$ in a weak magnetic field. The out-of-plane magnetic field opens up a non-trivial band gap at the 3-fold degenerate Dirac point, giving rise to nonzero thermal Hall signals that can be experimentally verified with transport measurements~\cite{Romhanyi2015}. The triplon band is not of the lowest energy and is only topologically nontrivial below the critical field. Since the triplons are bosonic, this band is only physically relevant with thermal excitations. These two factors could strongly suppress the strength of the thermal Hall effect in experiments. Moreover, at the theoretical level, the single-dimer model does not fully describe the nature of triplon excitations. Thus it is necessary to extend the tunable parameter space and the degrees of freedom to uncover more interesting physics. The pseudospin degrees of freedom from the non-equivalent dimers can also have nontrivial physical consequences by coupling to the external magnetic field~\cite{Mak2018}, and this is relatively unexplored in the literature. 
   
In this Letter, we construct a two-dimer model for frustrated Shastry-Sutherland models with the sublattice pseudospin degrees of freedom, that couples a tilted external magnetic field with both in-plane and out-of-plane components. Remarkably, the in-plane component gives rise to a new Dirac point between the lowest two triplon bands. The elementary excitations -- Dirac triplons (DT) -- near the Dirac point have nontrivial topological properties with $\pi$-Berry flux~\cite{Xu2016}. Thus an out-of-plane magnetic field opens up a gap at the Dirac point with nonzero Berry curvature hot spot around the Brillouin zone (BZ) center, or near the \emph{bottom} of the triplon band. In addition, the critical magnetic field increases with the tilt angle of the magnetic field. We explicitly compute the Hall response with different tilt angle and magnetic field strength, showing the pronounced tunable experimental signature for the thermal Hall effect (THE), and provide theoretical evidence for the existence of external field induced DT. We point out there is a sign change of the thermal Hall signals as a function of the titled angle, which is the typical feature of DT. The induced negative thermal Hall conductance reveals the flexible tunability of the thermal transport by the in-plane magnetic field. 

{\em Microscopic model.}--
At the microscopic level, the spin Hamiltonian with external magnetic field is given by~\cite{Romhanyi2015}:
\begin{equation}\label{eq:1}
\begin{aligned}[b]
H =&J\sum_{n.n.}\vec{S}_i\cdot\vec{S}_j+J'\sum_{n.n.n.}\vec{S}_i\cdot\vec{S}_j+\sum_{n.n.}\vec{D}_{ij}\cdot \left(\vec{S}_i\times\vec{S}_j\right)\\
&+\sum_{n.n.n.}\vec{D}'_{ij}\cdot \left(\vec{S}_i\times\vec{S}_j\right)+H_m.
\end{aligned}
\end{equation}
Where $J$ is the isotropic intra-dimer exchange, $\vec{D}_{ij}$ is the intra-dimer DM coupling, $J'$ is the inter-dimer exchange with $J'/J\approx0.63$, and $\vec{D}'_{ij}$ is the inter-dimer exchange. The triplon excitations come from the interactions between these spin dimers, and $H_m=-g\mu_B\sum_{i}\vec{h}\cdot\vec{S}_i$ is the Zeeman energy in the presence of an external magnetic field $\vec{h}=(h^x, h^y, h^z)$, with $g$ as the g-factor~\cite{Romhanyi2015,McClarty2017} and $\mu_B$ being the Bohr magneton.

We use the bond-operator formalism to represent the two spins $\vec{S}_1$ and $\vec{S}_2$ of each copper dimer~\cite{Sachdev1990}. The Hilbert space of an isolated dimer is spanned by a singlet $\ket{s}$ and three triplet states $\ket{t_x}$, $\ket{t_y}$, and $\ket{t_z}$. We introduce the singlet creation operator $s^{\dagger}$ and triplet creation operators $t^{\dagger}_{\alpha}~(\alpha= x, y, z)$~\cite{Sachdev1990} with the algebra $\left[s,s^{\dagger}\right]=1$, $\left[t_{\alpha},t^{\dagger}_{\beta}\right]=\delta_{\alpha\beta}$, and $\left[s,t^{\dagger}_{\alpha}\right]=0$, subject to the usual hard-core constraint $s^{\dagger}s+\sum_{\alpha}t^{\dagger}_{\alpha}t_{\alpha}=1$~\cite{Sachdev1990}. The spin operators of the dimer in bond-operator representation is given by $S_{1\alpha}+S_{2\alpha}=s^{\dagger}t_{\alpha}+t_{\alpha}^{\dagger}s$, $S_{1\alpha}-S_{2\alpha}=-i\sum_{\beta\gamma}\epsilon_{\alpha\beta\gamma}t^{\dagger}_{\beta}t_{\gamma}$, and $\vec{S}_1\cdot\vec{S}_2=\sum_{\beta}t^{\dagger}_{\beta}t_{\beta}-3/4$, where $\epsilon_{\alpha\beta\gamma}$ is the anti-symmetric Levi-Civita symbol. 

Using the bond-operator language, and a mean field approximation, we transform the spin Hamiltonian in (\ref{eq:1}) into an effective tight-binding (TB) model: $H=H_{site}+H_{hop}$. $H_{site}$ and $H_{hop}$ are the onsite energy terms for the two dimer and hopping matrix, respectively. The explicit form of the TB model is expressed as: 
\begin{equation}\label{eq:2}
\begin{aligned}[b]
H_{site}=&\sum_{\vec{r}}A^{\dagger}_{\vec{r}}\mathcal{N}A_{\vec{r}}+\sum_{\vec{r}'}B^{\dagger}_{\vec{r}'}\mathcal{N}B_{\vec{r}'},\\
H_{hop}=&\sum_{\vec{r}}\sum^4_n B^{\dagger}_{\vec{r}+\vec{\delta}_n}M(\vec{\delta}_n)A_{\vec{r}}+h.c.,
\end{aligned}
\end{equation}
where $A^{\dagger}_{\vec{r}}=(\tilde{t}^{\dagger}_{Ax,\vec{r}},~\tilde{t}^{\dagger}_{Ay,\vec{r}},~\tilde{t}^{\dagger}_{Az,\vec{r}})$ and $B^{\dagger}_{\vec{r}}=(\tilde{t}^{\dagger}_{Bx,\vec{r}},~\tilde{t}^{\dagger}_{By,\vec{r}},~\tilde{t}^{\dagger}_{Bz,\vec{r}})$ are the A and B dimers, and $\mathcal{N}$ is the diagonal onsite energy matrix. $\tilde{t}^{\dagger}_{x,\vec{r}}$ is the new triplon operator by rotating the Hilbert space. The coordinate $\vec{r}$ runs over the position of all unit cells, $\vec{\delta}_n$ is the nearest neighbour bond describing the two-dimer geometry as shown in Fig.~\ref{fig:1}(a), and $M$ are the hopping matrices between the dimers, with the following forms:
\begin{equation}\label{eq:3}
\begin{aligned}[b]
M(\pm \vec{\delta}_1)&=\frac{1}{2}
\renewcommand{\arraystretch}{1.2}
\begin{pmatrix}
0 & -D'_{\perp} & 0 \\
D'_{\perp} & 0& \pm\tilde{D}'_{\lVert}\\
0 & \mp\tilde{D}'_{\lVert} & 0
\end{pmatrix},\\
M(\pm \vec{\delta}_2)&=\frac{1}{2}
\renewcommand{\arraystretch}{1.2}
\begin{pmatrix}
0 & -D'_{\perp} & \mp\tilde{D}'_{\lVert} \\
D'_{\perp} & 0& 0\\
\pm\tilde{D}'_{\lVert} & 0& 0
\end{pmatrix}.
\end{aligned}
\end{equation}
where $D'_{\perp}$ is the out-of-plane component of inter-dimer DM vector, and $\tilde{D}'_{\lVert}=D'_{\lVert,s}-\frac{DJ'}{2J}$ is the effective in-plane DM component. We set  $J=1$, $\abs{\tilde{D}'_{\lVert}}/J=0.03$, and $D'_{\perp}/J=-0.03$ in accordance with previous studies~\cite{Romhanyi2015,McClarty2017}. 
The external magnetic field part in Eq.~(\ref{eq:1}) is given by:
\begin{equation}\label{eq:hm}
\begin{aligned}[b]
H_{m}=\sum_{\beta,\gamma,\vec{r}}ig\epsilon_{\alpha \beta \gamma}h_{\alpha}\tilde{t}^{\dagger}_{A\beta,\vec{r}}\tilde{t}_{A\gamma,\vec{r}}+\sum_{\beta,\gamma,\vec{r}'}ig\epsilon_{\alpha \beta \gamma}h_{\alpha}\tilde{t}^{\dagger}_{B\beta,\vec{r}'}\tilde{t}_{B\gamma,\vec{r}'}.
\end{aligned}
\end{equation}
This term modifies the onsite energy matrix $N$ with additional off-diagonal elements:
\begin{equation}
\begin{aligned}[b]
\renewcommand{\arraystretch}{1.2}
\mathcal{N}=
\begin{pmatrix}
J & igh_z & -igh_y \\
-igh_z & J& igh_x\\
igh_y & -igh_x & J
\end{pmatrix},
\end{aligned}
\end{equation}
which gives rise to exotic topological properties~\cite{Romhanyi2015,McClarty2017}. We fix the in-plane component of the magnetic field to be along the $x$ direction, let $\vec{h}=(h_i,0,h_z)$. 
To ensure the validity of the models in Eqs.~(\ref{eq:2}) and Eq.~(\ref{eq:hm}), the external magnetic field $\vec{h}$ has to be of the same order as the DM strength which is small compared to the exchange strengths $J$ and $J'$.
Thus, only terms linear in $\tilde{D}'_{\lVert}$, $D'_{\perp}$ and $\vec{h}$ are kept when transforming spin Hamiltonian in Eq.~(\ref{eq:1}) to the effective TB model.

\begin{figure}[t]
\centering  
\includegraphics[width=0.5\textwidth]{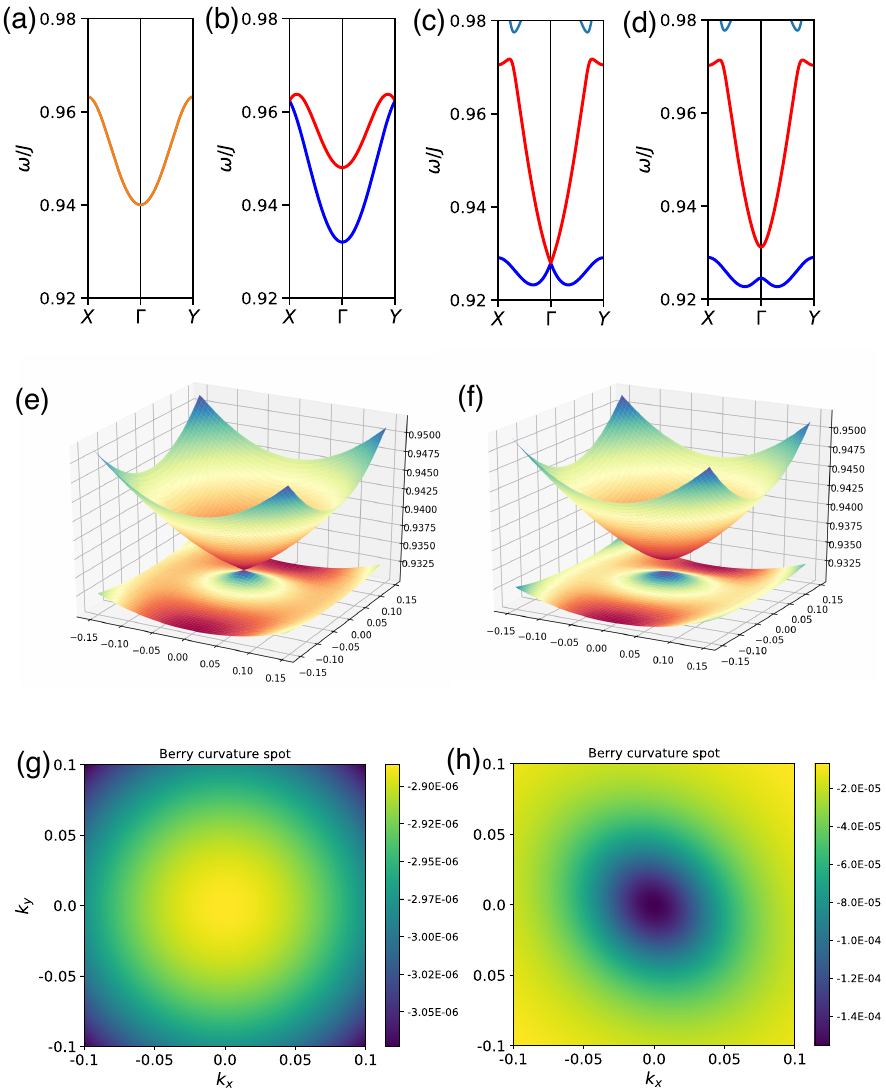}
\caption{\label{}
Band structure of triplon excitations in SSM. (a) Without magnetic field, the lowest triplon bands are 2-fold degenerate. (b) With only out-of-plane magnetic field $(h_z/J=0.008)$, the lowest triplon bands split into red and blue sub-bands. (c) With only in-plane component of magnetic field $(h_i/J=0.04)$, the lowest triplon bands also splits, and form Dirac point at $\Gamma$ point. (d) With both In-plane and out-of-plane magnetic field $(h_i/J=0.04,h_z/J=0.004)$, the Dirac point in (c) is gapped. The triplon band (e) near the $\Gamma$ point at $h_i/J=0.04$. (f) near the $\Gamma$ point at $h_i/J=0.04$ and $h_z/J=0.004$. (g) The Berry curvature hot spot with only out-of-plane magnetic field $(h_z/J=0.008)$ is nearly 0 around the BZ center. (h) With both in-plane and out-of-plane magnetic field, the Berry curvature hot spot has very large values, and gives rise to the thermal Hall effect.}
\label{fig:2}
\end{figure}

{\em Triplon band and $\pi$-flux Dirac Boson.}--
Starting from Eqs.~(\ref{eq:2}), the Hamiltonian of triplon excitations with pseudospin in momentum space can be written in the following form with the basis $(A_k,B_k)$:
\begin{equation}\label{eq:Dirac}
\begin{aligned}[b]
\mathcal{H}(k)=\sigma_x\otimes m_x+\sigma_y\otimes m_y +\sigma_0\otimes \mathcal{N},
\end{aligned}
\end{equation}
where $m_x(k)=\frac{\mathcal{M}(k)+\mathcal{M}^{\dagger}(k)}{2}$, and $m_y(k)=i\frac{\mathcal{M}^{\dagger}(k)-\mathcal{M}(k)}{2}$. $\mathcal{M}(k)=\sum^4_ne^{-i\vec{k}\cdot\vec{\delta}_n}M(\vec{\delta}_n)$ is a $3\cross3$ traceless matrix that can be obtained from Eq.~(\ref{eq:3}). $\sigma_x$ and $\sigma_y$ are Pauli matrices for the pseudospin, and $\sigma_0$ is $2\cross2$ identity matrix. The calculated band structures with and without magnetic fields are shown in Fig.~\ref{fig:2}. The original lowest bands in Fig.~\ref{fig:2}(a) show a parabolic feature at the BZ center are 2-fold degenerate. When out-of-plane magnetic field $h_z$ is turned on, the degeneracy of the lowest triplon bands is lifted. The gap $\Delta \omega$ of the excitation spectrum near the zone center is proportional to $h_z$, but vanishes at the middle point $X$ and $Y$ of the BZ edges. With only the in-plane magnetic fields, the lowest triplon bands undergo a Rashba-like splitting and form a gapless Dirac point at the $\Gamma$ point as shown in Fig.~\ref{fig:2}(c). The anisotropic dispersion of the DT is also shown in Fig.~\ref{fig:2}(e), which depends on the orientation angle of the in-plane component $\tan(\phi)=h_y/h_x$. Thus the direction of the in-plane magnetic field can be a novel tuning knob for the transport of DT in this quantum magnet system. 


We shall particularly focus on the case when the applied magnetic field has both non-zero in-plane and out-of-plane components. A finite $h_z$ component lifts the 2-fold degeneracy of the Dirac point and opens up a gap in the dispersion at the $\Gamma$ point as shown in Fig.~\ref{fig:2}(d)(f).  The nontrivial excitation gap results in a finite Berry curvature distribution around the zone center, which is near the bottom of the band as depicted in Fig.~\ref{fig:2}(h). In contrast, with only out-of-plane magnetic field ($h_z/J$=0.008), the Berry curvature is nearly zero at the band bottom  (Fig.~\ref{fig:2}(g)). 

To better understand the different roles of external magnetic components, we construct an effective low-energy theory of the lowest two bands near the $\Gamma$ point from the original TB Hamiltonian as follows:
\begin{equation}\label{eq:eff}
\begin{aligned}[b]
\mathcal{H}(k)=\frac{\hbar^2k^2}{2\mu}+\alpha h_i\left(\sigma_xk_x+\eta\sigma_yk_y\right) +\beta h_z\sigma_z,
\end{aligned}
\end{equation}
where $\mu$ is the effective mass of parabolic triplon band. $\alpha$, $\beta$ are constants determined by $D'_{\perp}$ and $\tilde{D}'_{\lVert}$, and $\eta$ is the coefficient for the anisotropic feature determined by the orientation of the in-plane component of magnetic field. This effective model exactly reproduces the band structure near the $\Gamma$ point, and reveals the low energy behavior of the DT excitations. 
One can immediately identify it with the Rashba spin-orbit coupling (SOC) Hamiltonian in the fermionic systems~\cite{Bercioux_2015, Manchon2015}. In our case, $h_i$ plays the role as the Rashba pseudospin-orbit coupling (POC) coefficient. The Dirac boson thus hosts a $\pi$ flux from the Berry phase of a closed circle around the Dirac point: $\oint_{C}\mel{\psi_k}{i\partial_k}{\psi_k}dl=-\pi$. When DT anti-crossing is gapped by the out-of-plane magnetic field $h_z$, this point-like flux $\phi$ gives rise to the nonzero Berry curvature hot spot near the $\Gamma$ point. 

\begin{figure}[t]
\centering  
\includegraphics[width=0.48\textwidth]{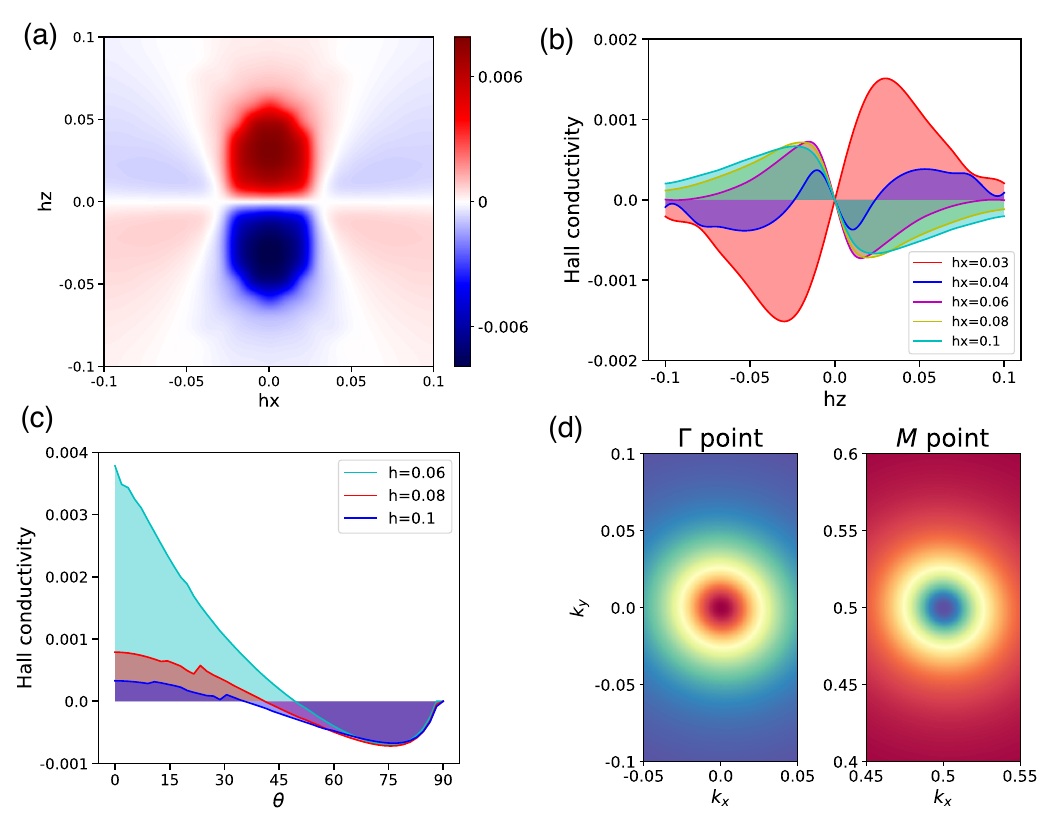}
\caption{\label{}
The thermal Hall conductivity at 10 K. (a) The diagram of thermal Hall conductivity is a function of the magnetic field in $h_x-h_z$ plane. (b) Thermal Hall conductivity along the $h_z$ axis with different magnitudes of $h_x$ component. (c) The angle-dependent Hall conductivity under a tilted magnetic field with different magnitudes. (d) The Berry curvature distributions of lowest triplon band. The left panel shows the hotspot around $\Gamma$ point (BZ center), and right panel shows the hotspot distribution around M point (BZ corner) with opposite sign.}
\label{fig:3}
\end{figure}

{\em Bosonic thermal Hall effect.}--
Given that the triplons are charge neutral, the thermal Hall effect (THE) allows us to probe the Berry curvature distribution in the bosonic systems. The thermal Hall conductivity $\kappa^{xy}$ is given as follows~\cite{Romhanyi2015,Matsumoto2011}: 
\begin{equation}\label{eq:the}
\begin{aligned}[b]
\kappa^{xy}=\frac{k_{B}^2T}{(2\pi)^2\hbar}\sum_{n}\int d^2k c_{2}(\rho_n)\Omega^{xy}_n(\vec{k}),
\end{aligned}
\end{equation}   
where $k_B$ is Boltzmann constant, $\rho_{n}=\frac{1}{e^{\omega_n\beta}-1}$ is the Bose-Einstein (BE) distribution function, $\beta=1/(k_BT)$, and $c_2(u)$ is given by~\cite{Matsumoto2011}: 
\begin{equation}\label{eq:}
\begin{aligned}[b]
c_2(u)=(1+u)\ln^2\left(1+\frac{1}{u}\right)-\ln^2(u)-2Li_2(-u),
\end{aligned}
\end{equation} 
with $Li_2(u)=\sum^{\infty}_{k=1}\frac{u^k}{k^2}$ as the polylogarithm function, and $\Omega^{xy}_n(k_x,k_y)$ is the Berry curvature. 
In the low-temperature limit, the dominant contribution to $\kappa^{xy}$ comes from the lower bands, and the non-zero Berry curvature shown in Fig.~\ref{fig:2}(h)  contributes to the thermal Hall conductivity. Thus, the change of the total thermal Hall conductivity provides a way to probe this $\pi$-flux boson in the triplon system. Figure~\ref{fig:3}(a) shows the calculated Hall conductivity in the  $h_x-h_z$ parameter space. The Hall signal is symmetric about $h_x=0$. It reaches extremal value at $h_z=0.03$, and approaches zero as $h_x$ increases. This is can be seen clearly in Fig.~\ref{fig:3}(b). When $h_x>0.04$, the Hall signal becomes negative at a positive $h_z$ and positive at a negative $h_z$. This interesting feature is due to the fact that the Berry curvature around the gapped DT point is always $opposite$ to those around the $X/Y$ point when the gap is opened by the $h_z$ field. To verify this, we plot the schematic distributions of Berry curvature at the $\Gamma$ and M point the lowest triplon band at the $h_x=0.02$ and $h_z=0.001$, which is shown in Fig.~\ref{fig:3}(d). The Berry curvature in the zone center ($\Gamma$ point) shows a negative value, while a positive Berry curvature develops at the zone corner (M point). 

The switch of the sign of the thermal Hall signal can be easily probed with an angle-dependent Hall measurement, which provides a concrete test for the existence of magnetic field-induced Dirac triplon excitations. As shown in Fig.~\ref{fig:3}(c), the angle-dependent Hall signal with different magnetic field strength are plotted, where $\theta=\arctan(h_x/h_z)$ is the angle between tilted magnetic field $\vec{h}$ and the vertical z-axis. The Hall signals are large at a very small $h_x/h_z$ ratio, and decay fast by increasing the ratio when $\abs{\vec{h}}<0.08$ (red line). When the threshold magnetic field strength $h_c\approx1.4$ Tesla is reached, the topological nature of the triplon bands is lost and the Hall signal $\kappa^{xy}$ is suppressed. One can see the angle-dependent Hall signal (purple line with $\abs{h}=0.1$) is almost 0 at very small angle, where the band is trivial when $\theta < 30^{\operatorname{\omicron}}$. However, the situation is quite different when tuning the tilted angle larger than a critical value. The Hall signal shows a sign change and reaches a maximal negative value around the angle $77^{\operatorname{\omicron}}$. This sign change could be the solid evidence for the gapped $\pi$-flux DT induced by the in-plane magnetic field. 

\begin{figure}[t]
\centering  
\includegraphics[width=0.45\textwidth]{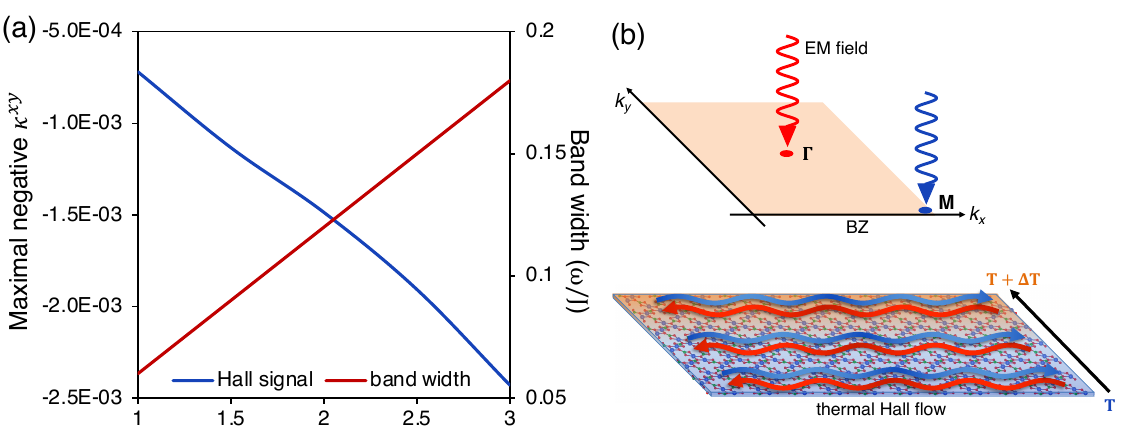}
\caption{\label{}
(a) Enhancement of Hall signal by changing strength DM interactions. The x axis is the strength of DM term from $D'_{\perp}$ to $3D'_{\perp}$. Both bandwidth and maximal negative Hall signals are plot. (b) The schematic plot of triplon density amplification by external electromagnetic field at different k points, which shows the direction-tunable features of the thermal Hall flow.}
\label{fig:4}
\end{figure}
 
The thermal Hall signal in $\rm SrCu_2(BO_3)_2$ is small due to the  weak DM interaction, making the experimental detection more challenging. It has been reported that the proximity effect is critical for the promotion of the DM interactions by increasing the orbital hybridization at the interface, which may provide a potential approach for $\rm SrCu_2(BO_3)_2$ shown in Fig.~\ref{fig:1}(b). In order to maintain the validity of the effective TB model, we limit the enhancement of the DM parameter $\tilde{D}'_{\lVert}$ and $D'_{\perp}$ to be within three times the current values. Since the bandwidth is proportional to the DM interactions, the large DM interaction will give rise to a large triplon bandwidth. In addition, since the DM interactions mix the singlet ground state with three triplet excitations, increasing of DM interactions will decrease the excitation gap, which gives rise to larger triplon populations. Both will lead to more pronounced triplon thermal Hall signal.

Numerical calculations show the bandwidth without magnetic field is linear in the DM interactions as plotted in Fig.~\ref{fig:4}(a). The analytic solutions of band width can also be found by solving the Hamiltonian $\mathcal{H}(k)$ at $k=0$ without the last term in Eq.~(\ref{eq:Dirac}), which gives the bandwidth $\Delta\omega=2D'_{\perp}$. We calculate the maximal negative Hall conductivity induced by gapped DT by searching the $h_x-h_z$ plane, which also shows the surprising linear behaviours with respect to the increasing of $D'_{\perp} (\tilde{D}'_{\lVert})$. One can roughly understand this linear relation from the change of the activation gap $\Delta E=J-2D'_{\perp}$ between the singlet and triplet excitations. The total population is approximately $P_t=e^{-\Delta E/k_BT}\approx 1-\beta\Delta E$. The maximal negative Hall conductivity mainly dominated by the gapped DT in the lowest triplon bands is $\kappa^{xy}_N\approx P_t \Omega^{xy}(0)S_\Gamma$, where $\Omega^{xy}(0)S_\Gamma$ gives the $-\pi$ flux. Thus we have $\kappa^{xy}_N\approx-\pi(1-\beta J+2\beta D'_{\perp})$. With increasing the $D'_{\perp} (\tilde{D}'_{\lVert})$, the maximal negative Hall conductivity also follows the linear relation.  

Triplon amplification by driving field is another way to enhance the Hall effect by the parametric instability while preserving the the magnetic order in bulk state~\cite{Malz2019}. With a coherent driving electromagnetic (EM) field, the anomalous triplon pairing terms can be created: 
\begin{equation}\label{eq:the}
\begin{aligned}[b]
H_{int}=\sum_{k}\frac{g_k}{2}(t^{\dagger}_{-k}t^{\dagger}_k b+b^{\dagger}t_{k}t_{-k}),
\end{aligned}
\end{equation} 
where the field operator $b\approx\beta\exp(-i\Omega_0t)$ is the external EM field or pump photon, and $g_k\approx g$ is the coupling strength between external field and triplon excitation that can be treat as momentum independent in a small bandwidth regime. The triplon amplification requires several conditions. First, there should be the conservation of momentum of the triplon pair. Second, the energy of the triplon pair should match the driving pump field. The last important condition is that the strength of coupling has to exceed detuning and damping in order to have a triplon density accumulation. The triplon density is determined by the time-dependent equation of motion as follows~\cite{Malz2019}:
\begin{equation}\label{eq:the}
\begin{aligned}[b]
i\frac{d T_k(t)}{dt}=\tilde{\Omega}_k T_k(t),
\end{aligned}
\end{equation} 
where $T_k(t)=(\langle t_k \rangle, \langle t^{\dagger}_{-k} \rangle)$ is the classical density of the triplon fields, and $\tilde{\Omega}_k$ is the dynamical matrix, which has the eigenvalues:
\begin{equation}\label{eq:the}
\begin{aligned}[b]
\omega_{k,\pm}=\frac{\omega_k-\omega_{-k}}{2}-\frac{i\gamma}{2}\pm\sqrt{\frac{(\omega_k+\omega_{-k}-\Omega_0)^2}{2}-\epsilon^2},
\end{aligned}
\end{equation} 
where $\gamma$ gives the dissipative damping, and $\epsilon=g\beta$ is the overall coupling strength. When the detuning term $\omega_k+\omega_{-k}-\Omega_0\approx0$, the imaginary part of $\omega_{k,+}$ becomes $\epsilon-\gamma/2$, and the triplon density at $k$ is $\langle t_k \rangle \propto e^{(\epsilon-\gamma/2)t}$. When the coupling strength $\epsilon$ exceeds the dissipation $\gamma$, there is an exponential growth of the triplon density of mode $k$, which means a resonant amplification is achieved. Due to the linear band dispersions $\omega=\omega_{\Gamma}+v_{\Gamma} k+O(k^2)$  with slope $v_{\Gamma}$ of DT, the momentum and energy matching is easily fulfilled. We can estimate the resonant driving frequency $\Omega_0=2\omega_{\Gamma}$ around the DT only with the linear dispersion. We use the exchange parameter $J=722$ GHz in $\rm SrCu_2(BO_3)_2$ system, and it gives rise to the resonant frequency $\Omega_0=1.34$ THz. SSM also holds a spin-1 Dirac cone at the BZ corner with a relatively higher energy level, which requires a higher resonant frequency of about $1.44$ THz to achieve the amplification. Thus, one can realise the triplon density amplification at selected momenta by choosing the driving light with different frequencies. As a result, the negative thermal Hall conductance induced by DT can be potentially enhanced by the triplon density amplification. Additionally, one can achieve directional tunability of thermal Hall flow by precisely controlling the EM field at different k points, i.e., the amplified triplons near the $\Gamma$ point with non-zero Berry curvature will hold the flow to the left, while the amplified triplons at the $M$ point will hold the flow in the opposite direction because of the opposite Berry curvature, as shown in the schematic plot Fig.~\ref{fig:4}(b). 

{\em Conclusions.}---To summarize, we found that the in-plane magnetic field ($h_x$) gives rise to a new type of Dirac point in the lowest bands at the BZ centre (momentum $k=0$) of the SSM, which has a $\pi$ Berry flux nature. The out-of-plane component $h_z$ thus opens up a gap with finite Berry curvature that is opposite to the curvature at the band edge. The induced negative thermal Hall conductance shows the possibility of the tunable nontrivial thermal transport. We also show that the negative thermal Hall conductance can be potentially enhanced by increasing the DM interaction and using an EM field-driven amplification. Based on the TB model, we develop an effective Low-energy model showing that $h_x$ induces a Rashba-like SOC, leading to useful insight into the bosonic Rashba SOC physics. One can possibly detect this SOC induced pseudo-spin texture distributions in momentum space and pseudo-spin density. It is also worth noting that the tilted magnetic field induces flat lowest bands, and the bandwidths are reduced by about 75\% along certain directions, which may hold the strongly correlated bosons. Once these interaction effects are added, bosonic systems hold the promise of realising new interaction-driven topological phases, where magnetic-induced flat triplon bands are ideally suited for realising the complex bosonic phases in a controllable manner.

\begin{acknowledgements}
Y.B would like to acknowledge the support by the National Research Foundation, Singapore under the NRF fellowship award (NRF-NRFF12-2020-005). SH and DN would like to acknowledge the support by Ministry of Education, Singapore, under grant AcRF TIER 1 2019-T1-002-050 (RG 148/19 (S)). PS acknowledges financial support from the Ministry of Education, Singapore through MOE2018-T1-001-021.
\end{acknowledgements}

\bibliography{triplon_bib_2}

\onecolumngrid
\clearpage
\begin{center}
\textbf{\large Supplemental Materials: Negative Thermal Hall Conductance in Two-Dimer Shastry-Sutherland Model with $\pi$-flux Dirac Triplon}
\end{center}
\setcounter{equation}{0}
\setcounter{figure}{0}
\setcounter{table}{0}
\setcounter{page}{1}
\makeatletter
\renewcommand{\theequation}{S\arabic{equation}}
\renewcommand{\thefigure}{S\arabic{figure}}
\renewcommand{\bibnumfmt}[1]{[S#1]}
\renewcommand{\citenumfont}[1]{S#1}

\section{Bond-operator representation of spin dimers}\label{A}
We use the Einstein summation convention in the expressions, and therefore, the repeated indices are implicitly summed over for achieving notational brevity. The spin operators $\vec{S}_1$ and $\vec{S}_2$ in the dimer have the SU(2) algebra:
\begin{eqnarray}
&&[S_{1\alpha},S_{1\beta}]=i\epsilon_{\alpha\beta\gamma}S_{1\gamma},\\
&&[S_{2\alpha},S_{2\beta}]=i\epsilon_{\alpha\beta\gamma}S_{2\gamma}, \\
&&[S_{1\alpha},S_{2\beta}]=0.
\end{eqnarray}
Spin operator can be rewritten as :
\begin{equation}
\vec{S}_i = \frac{1}{2}\sum_{\mu\nu}c^{\dagger}_{i\mu}\vec{\sigma}_{\mu\nu}c_{i\nu},
\end{equation}
where $c^{\dagger}_{i\mu}$ ($c_{i\mu}$) is electron creation (annihilation) operator, which creates (annihilates) single-particle state with spin component $\mu\in(1,2)$ on the site $i$:  
\begin{eqnarray}
&&c^{\dagger}_{i1}\ket{0} =\ket{\uparrow},\qquad c^{\dagger}_{i2}\ket{0} =\ket{\downarrow},\\
&&c_{i1}\ket{\uparrow} =\ket{0},\qquad c_{i2}\ket{\downarrow} =\ket{0}.
\end{eqnarray}
Now we can represent spin operator by bond-operator method. Consider the nonzero matrix elements ($\mel{t_{\alpha}}{S_{1\alpha}}{t_{\beta}}$) are: 
\begin{eqnarray}
&&\mel{s}{S_{1\alpha}}{t_\beta}=\frac{1}{2}\delta_{\alpha\beta},\\
&&\mel{t_\beta}{S_{1\alpha}}{s}=\frac{1}{2}\delta_{\alpha\beta}, \\
&&\mel{t_{\alpha}}{S_{1\beta}}{t_{\gamma}}=\frac{-i}{2}\epsilon_{\alpha\beta\gamma},
\end{eqnarray}
we obtain:
\begin{eqnarray}
&&S_{1\alpha}=\frac{1}{2}\left(s^{\dagger}t_{\alpha}+t_{\alpha}^{\dagger}s-i\epsilon_{\alpha\beta\gamma}t^{\dagger}_{\beta}t_{\gamma}\right),\\
&&S_{2\alpha}=\frac{-1}{2}\left(s^{\dagger}t_{\alpha}+t_{\alpha}^{\dagger}s+i\epsilon_{\alpha\beta\gamma}t^{\dagger}_{\beta}t_{\gamma}\right).
\end{eqnarray}
In the dimer with two spins, we have the relations:
\begin{eqnarray}
&&S_{1\alpha}+S_{2\alpha}=-i\epsilon_{\alpha\beta\gamma}t^{\dagger}_{\beta}t_{\gamma},\\
&&S_{1\alpha}-S_{2\alpha}=s^{\dagger}t_{\alpha}+t^{\dagger}_{\alpha}s.
\end{eqnarray}
The product $\vec{S}_1\cdot\vec{S}_2$ in bond-operator representation can be written as:
\begin{equation}\label{eq:A18}
\begin{aligned}[b]
\vec{S}_1\cdot\vec{S}_2=&\frac{1}{2}\left[\left(\vec{S}_1+\vec{S}_2\right)^2-\vec{S}^2_1-\vec{S}^2_2\right]=\frac{-1}{2}\epsilon_{\alpha\beta\gamma}\epsilon_{\alpha\mu\nu}t^{\dagger}_{\beta}t_{\gamma}t^{\dagger}_{\mu}t_{\nu}-\frac{3}{4}=\frac{1}{2}\left(t^{\dagger}_{\beta}t_{\gamma}t^{\dagger}_{\gamma}t_{\beta}-t^{\dagger}_{\beta}t_{\gamma}t^{\dagger}_{\beta}t_{\gamma}\right)-\frac{3}{4}\\
=&t^{\dagger}_{\beta}t_{\beta}-\frac{3}{4},
\end{aligned}
\end{equation}
where 
\begin{equation}
\begin{aligned}[b]
t^{\dagger}_{\beta}t_{\gamma}t^{\dagger}_{\gamma}t_{\beta}=\ketbra{t_i}t^{\dagger}_{\beta}t_{\gamma}t^{\dagger}_{\gamma}t_{\beta}\ketbra{t_j}=2\ketbra{t_\beta}=2t^{\dagger}_{\beta}t_{\beta},
\end{aligned}
\end{equation}
and
\begin{equation}
t^{\dagger}_{\beta}t_{\gamma}t^{\dagger}_{\beta}t_{\gamma}=\ketbra{t_i}t^{\dagger}_{\beta}t_{\gamma}t^{\dagger}_{\beta}t_{\gamma}\ketbra{t_j}=0
\end{equation}
for $\beta \neq \gamma$.

The Hamiltonian of intra-dimer is:
\begin{equation}\label{eq:A21}
\begin{aligned}[b]
H =J\vec{S}_1\cdot\vec{S}_2+\vec{D}_{12}\cdot \left(\vec{S}_1\times\vec{S}_2\right)
=JS_{1\alpha}S_{2\alpha}+\epsilon_{\alpha\beta\gamma}D^{\alpha}_{12}S_{1\beta}S_{2\gamma},
\end{aligned}
\end{equation} 
where $J$ is the exchange coupling, $\vec{D}_{12}$ is the intra-dimer Dzyaloshinskii-Moriya interaction, and
\begin{equation}\label{eq:A22}
\begin{aligned}[b]
4S_{1\beta}S_{2\gamma}=&\left(s^{\dagger}t_{\beta}+t_{\beta}^{\dagger}s-i\epsilon_{\beta\mu\nu}t^{\dagger}_{\mu}t_{\nu}\right)\left(-s^{\dagger}t_{\gamma}-t_{\gamma}^{\dagger}s-i\epsilon_{\gamma\rho\sigma}t^{\dagger}_{\rho}t_{\sigma}\right)\\
=&-s^{\dagger}t_{\beta}s^{\dagger}t_{\gamma}-\underline{s^{\dagger}t_{\beta}t_{\gamma}^{\dagger}s}-\underline{i\epsilon_{\gamma\rho\sigma}s^{\dagger}t_{\beta}t^{\dagger}_{\rho}t_{\sigma}}-\underline{t_{\beta}^{\dagger}ss^{\dagger}t_{\gamma}}-t_{\beta}^{\dagger}st_{\gamma}^{\dagger}s-i\epsilon_{\gamma\rho\sigma}t_{\beta}^{\dagger}st^{\dagger}_{\rho}t_{\sigma}+i\epsilon_{\beta\mu\nu}t^{\dagger}_{\mu}t_{\nu}s^{\dagger}t_{\gamma}+\underline{i\epsilon_{\beta\mu\nu}t^{\dagger}_{\mu}t_{\nu}t_{\gamma}^{\dagger}s}-\underline{\epsilon_{\beta\mu\nu}\epsilon_{\gamma\rho\sigma}t^{\dagger}_{\mu}t_{\nu}t^{\dagger}_{\rho}t_{\sigma}}\\
=&\delta_{\beta\gamma}\left(s^{\dagger}s+t^{\dagger}_{\sigma}t_{\sigma}\right)+i\epsilon_{\sigma\beta\gamma}s^{\dagger}t_{\sigma}-i\epsilon_{\sigma\beta\gamma}t^{\dagger}_{\sigma}s-t^{\dagger}_{\beta}t_{\gamma}-t^{\dagger}_{\gamma}t_{\beta}\\
=&\delta_{\beta\gamma}+i\epsilon_{\sigma\beta\gamma}s^{\dagger}t_{\sigma}-i\epsilon_{\sigma\beta\gamma}t^{\dagger}_{\sigma}s-t^{\dagger}_{\beta}t_{\gamma}-t^{\dagger}_{\gamma}t_{\beta}.
\end{aligned}
\end{equation} 
Inserting Eqs.~(\ref{eq:A18}) and (\ref{eq:A22}), Eq.~(\ref{eq:A21}) has the form:
\begin{equation}\label{eq:A23}
\begin{aligned}[b]
H=&JS_{1\alpha}S_{2\alpha}+\epsilon_{\alpha\beta\gamma}D^{\alpha}_{12}S_{1\beta}S_{2\gamma}\\
=&Jt^{\dagger}_{\alpha}t_{\alpha}-\frac{3}{4}J+\frac{iD^{\alpha}_{12}}{2}s^{\dagger}t_{\alpha}-\frac{iD^{\alpha}_{12}}{2}t^{\dagger}_{\alpha}s\\
=&T^{\dagger}NT,
\end{aligned}
\end{equation} 
where $T^{\dagger}=\left(s^{\dagger},~t^{\dagger}_x,~t^{\dagger}_y,~t^{\dagger}_z\right)$, and
\begin{equation}
\begin{aligned}[b]
\renewcommand{\arraystretch}{2}
N=
\begin{pmatrix}
-\dfrac{3}{4}J & \dfrac{iD^{x}_{12}}{2} & \dfrac{iD^{y}_{12}}{2} & \dfrac{iD^{z}_{12}}{2}\\
-\dfrac{iD^{x}_{12}}{2} & \dfrac{1}{4}J & 0& 0\\
-\dfrac{iD^{y}_{12}}{2} & 0 & \dfrac{1}{4}J & 0\\
-\dfrac{iD^{z}_{12}}{2} & 0 & 0 & \dfrac{1}{4}J
\end{pmatrix}.
\end{aligned}
\end{equation}
From the above repressions, one can find that the intra-dimer DM interactions mix singlet and triplet states. And the new energy levels of spin dimer are rearranged as:
\begin{eqnarray}
&&E_1=\frac{1}{4}\left(-J-2\sqrt{\abs{D_{12}}^2+ J^2}\right),\qquad E_2=\frac{J}{4},\\
&&E_3=\frac{1}{4}\left(-J+2\sqrt{\abs{D_{12}}^2+ J^2}\right),\qquad E_4=\frac{J}{4}.
\end{eqnarray}
where $D_{12}=\abs{\vec{D}_{12}}$. By dropping the constant term $\frac{-3J}{4}$ and neglecting the high-order terms with $\frac{(D^{
\alpha}_{12})^2}{J^2}$, the Hamiltonian in ~(\ref{eq:A23}) is only with diagonal terms, and can be written as:
\begin{equation}\label{eq:}
\begin{aligned}[b]
H
\approx J\left(t^{\dagger}_{\alpha}+\frac{iD^{\alpha}_{12}}{2J}s^{\dagger}\right)\left(t_{\alpha}-\frac{iD^{\alpha}_{12}}{2J}s\right)=\tilde{T}^{\dagger}N\tilde{T},
\end{aligned}
\end{equation} 
where $\tilde{T}$ is the 3-component basis vector
\begin{equation}\label{eq:31}
\begin{aligned}[b]
\tilde{T}=
\begin{pmatrix}
\tilde{t}_x\\
\tilde{t}_y\\
\tilde{t}_z
\end{pmatrix}
=
\renewcommand{\arraystretch}{2}
\begin{pmatrix}
t_x-\dfrac{iD^{x}_{12}}{2J}s\\
t_y-\dfrac{iD^{y}_{12}}{2J}s\\
t_z-\dfrac{iD^{z}_{12}}{2J}s
\end{pmatrix},
\end{aligned}
\end{equation} 
and $N$ is the diagonal matrix:
\begin{equation}\label{eq:}
\begin{aligned}[b]
N=
\begin{pmatrix}
J&0&0\\
0&J&0\\
0&0&J
\end{pmatrix}.
\end{aligned}
\end{equation} 
The Zeeman terms in Hamiltonian:
\begin{equation}\label{eq:}
\begin{aligned}[b]
-g_zh^z\sum_{i}S^z_i=&-g_zh^z\left(S^z_{1}+S^z_{2}\right)\\
=&ig_zh^z\left(t_{x}^{\dagger}t_{y}-t_{y}^{\dagger}t_{x}\right)\\
\approx&ig_zh^z\Bigl[\left(t_{x}^{\dagger}+\frac{iD^{x}_{12}}{2J}s^{\dagger}\right)\left(t_{y}-\frac{iD^{y}_{12}}{2J}s\right)-\left(t_{y}^{\dagger}+\frac{iD^{y}_{12}}{2J}s^{\dagger}\right)\left(t_{x}-\frac{iD^{x}_{12}}{2J}s\right)\Bigr]\\
=&ig_zh^z\left(\tilde{t}_{x}^{\dagger}\tilde{t}_{y}-\tilde{t}_{y}^{\dagger}\tilde{t}_{x}\right).
\end{aligned}
\end{equation} 
We write the above terms in the new basis vector~(\ref{eq:31}) by only retaining terms up to linear order of $h^z$, since the magnetic field $h^z$ has the same magnitude with DM strength $D$ and $D'$, which are small compared with the exchange strength $J$ and $J'$.

\section{Effective Hamiltonian with pseudo-spin freedom}\label{B}
The effective Hamiltonian can be derived by the 2-dimer TB model. The full $6\times6$ Hamiltonian is written as:
\begin{equation}\label{eq:hk}
\begin{aligned}[b]
\renewcommand{\arraystretch}{2}
\mathcal{H}(k)=
\begin{pmatrix}
J&igh_z&-igh_y&0&D'_{\perp}\gamma_3&i\tilde{D}'_{\lVert}\gamma_2\\
-igh_z&J&igh_x&-D'_{\perp}\gamma_3&0&-i\tilde{D}'_{\lVert}\gamma_1\\
igh_y&-igh_x&J&-i\tilde{D}'_{\lVert}\gamma_2&i\tilde{D}'_{\lVert}\gamma_1&0\\
0&-D'_{\perp}\gamma_3&i\tilde{D}'_{\lVert}\gamma_2 &J&igh_z&-igh_y\\
D'_{\perp}\gamma_3&0 &-i\tilde{D}'_{\lVert}\gamma_1&-igh_z&J&igh_x\\
-i\tilde{D}'_{\lVert}\gamma_2 &i\tilde{D}'_{\lVert}\gamma_1&0 &igh_y&-igh_x&J
\end{pmatrix},
\end{aligned}
\end{equation}
where $\gamma_1= sin(\vec{k}\cdot\vec{\delta}_1)$, $\gamma_2=sin(\vec{k}\cdot\vec{\delta}_2)$, and $\gamma_3=cos(\vec{k}\cdot\vec{\delta}_1)+cos(\vec{k}\cdot\vec{\delta}_2)$. The eigenvalues are:
\begin{equation}\label{eq:eff}
\begin{aligned}[b]
\omega_1(k)=J\pm\sqrt{J^2+Root(f(k),1)},\\
\omega_2(k)=J\pm\sqrt{J^2+Root(f(k),2)},\\
\omega_3(k)=J\pm\sqrt{J^2+Root(f(k),3)},
\end{aligned}
\end{equation}
where f(k) is a 3-order polynomial, which has 3 roots labeled by $Root(f(k),n)$. Without magnetic field ($\vec{h}=0$), the eigenvalues of $6\times6$ Hamiltonian are:
\begin{equation}\label{eq:eff}
\begin{aligned}[b]
\omega_1(k)&=J-\sqrt{(D'_{\perp})^2+(\tilde{D}'_{\lVert})^2+\frac{1}{2}((D'_{\perp})^2-(\tilde{D}'_{\lVert})^2)(cos(2k_x)+cos(2k_y))+2(D'_{\perp})^2 cos(k_x)cos(k_y)},\\
\omega_2(k)&=J-\sqrt{(D'_{\perp})^2+(\tilde{D}'_{\lVert})^2+\frac{1}{2}((D'_{\perp})^2-(\tilde{D}'_{\lVert})^2)(cos(2k_x)+cos(2k_y))+2(D'_{\perp})^2 cos(k_x)cos(k_y)},\\
\omega_3(k)&=J,\\
\omega_4(k)&=J,\\
\omega_5(k)&=J+\sqrt{(D'_{\perp})^2+(\tilde{D}'_{\lVert})^2+\frac{1}{2}((D'_{\perp})^2-(\tilde{D}'_{\lVert})^2)(cos(2k_x)+cos(2k_y))+2(D'_{\perp})^2 cos(k_x)cos(k_y)},\\
\omega_6(k)&=J+\sqrt{(D'_{\perp})^2+(\tilde{D}'_{\lVert})^2+\frac{1}{2}((D'_{\perp})^2-(\tilde{D}'_{\lVert})^2)(cos(2k_x)+cos(2k_y))+2(D'_{\perp})^2 cos(k_x)cos(k_y)},
\end{aligned}
\end{equation}
One can see that the 2-fold degeneracy of each band due to the sublattice pseudo-spin. We denote the $\ket{\phi_1(k)}$ and $\ket{\phi_2(k)}$ are the eigenstates of lowest 2-fold degenerate triplon bands, thus the sub-Hamiltonian of lowest bands can be constructed as:
\begin{equation}\label{eq:}
\begin{aligned}[b]
\mathcal{H}_{eff}=
\begin{pmatrix}
\mel{\phi_1(k)}{\mathcal{H}(k)}{\phi_1(k)}&\mel{\phi_1(k)}{\mathcal{H}(k)}{\phi_2(k)}\\
\mel{\phi_2(k)}{\mathcal{H}(k)}{\phi_1(k)}&\mel{\phi_2(k)}{\mathcal{H}(k)}{\phi_2(k)}
\end{pmatrix}=
\begin{pmatrix}
\frac{k^2}{m}+\mel{\phi_1(k)}{\mathcal{H}_{h_z}}{\phi_1(k)}&\mel{\phi_1(k)}{\mathcal{H}_{h_x}}{\phi_2(k)}\\
\mel{\phi_2(k)}{\mathcal{H}_{h_x}}{\phi_1(k)}&\frac{k^2}{m}+\mel{\phi_2(k)}{\mathcal{H}_{h_z}}{\phi_2(k)}
\end{pmatrix},
\end{aligned}
\end{equation} 
where $\frac{k^2}{m}$ term shows the parabolic feature of the band bottom, $\mathcal{H}_{h_z}$ is the matrix in (\ref{eq:hk}) only with $h_z$ terms, and $\mathcal{H}_{h_x}$ is the matrix only with $h_x$ terms.

\section{Rotation wave approximations for triplon amplification}\label{C}
The triplon Hamiltonian can be rewritten without considering the driving field:
\begin{equation}\label{eq:}
\begin{aligned}[b]
H_0=\sum_k \omega_k \ketbra{t_k},
\end{aligned}
\end{equation} 
where $\ket{t_k}=t^{\dagger}_k\ket{0}$, $\ket{0}=t_k\ket{t_k}$. Using the replacement $t^{\dagger}_{k}t_k\to\ketbra{t_k}$ (spectral theory), the interacting Hamiltonian now is:
\begin{equation}\label{eq:}
\begin{aligned}[b]
H_{int}&=\sum_{k}\left(\frac{g_k}{2}t^{\dagger}_{-k}t^{\dagger}_k b+\frac{g_k}{2}^*b^{\dagger}t_{k}t_{-k}\right)\\
&=\sum_{k}\left(\frac{G_k}{2}b\ketbra{t_{-k}}{0}+\frac{G_k}{2}^*b^{\dagger}\ketbra{0}{t_{-k}}\right),
\end{aligned}
\end{equation}
Replacing the $b$ field by the classical one $b(t)=\beta e^{-i\Omega_0 t}+\beta^* e^{i\Omega_0 t}$, one have:
\begin{equation}\label{eq:}
\begin{aligned}[b]
H_{int}=\sum_{k}(Re^{-i\Omega_0 t}+\tilde{R}e^{i\Omega_0 t})\ketbra{t_{-k}}{0}
+\sum_{k}(\tilde{R}^*e^{-i\Omega_0 t}+R^*e^{i\Omega_0 t})\ketbra{0}{t_{-k}},
\end{aligned}
\end{equation}
where $R=\frac{G_k\beta}{2}$, and $\tilde{R}=\frac{G_k\beta^*}{2}$. We can remove the time-dependence by using the time translation operator $U=e^{iht}$ with $h=\sum_k \omega_0\ketbra{t_k}$.  We can write the time translation operator as:
\begin{equation}\label{eq:}
\begin{aligned}[b]
e^{iht}&=e^{i\omega_0t\sum_k\ketbra{t_k}}=\sum_n\frac{(i\omega_0t)^n}{n!}(\sum_k\ketbra{t_k})^n\\
&=\sum_{n=1}\frac{(i\omega_0t)^n}{n!}(\sum_k\ketbra{t_k})+1\\
&=e^{i\omega_0t}\sum_k\ketbra{t_k} +\ketbra{0}, \\
e^{-iht}&=e^{-i\omega_0t}\sum_k\ketbra{t_k} +\ketbra{0}.
\end{aligned}
\end{equation}
Now the interacting Hamiltonian is:
\begin{equation}\label{eq:}
\begin{aligned}[b]
e^{iht}H_{int}e^{-iht}=&\sum_{k}(Re^{-i(\Omega_0-\omega_0) t}+\tilde{R}e^{i(\Omega_0+\omega_0) t})\ketbra{t_{-k}}{0}+\sum_{k}(\tilde{R}^*e^{-i(\Omega_0+\omega_0) t}+R^*e^{i(\Omega_0-\omega_0) t})\ketbra{0}{t_{-k}}\\
\approx&\sum_{k}Re^{-i(\Omega_0-\omega_0) t}\ketbra{t_{-k}}{0}+\sum_{k}R^*e^{i(\Omega_0-\omega_0) t}\ketbra{0}{t_{-k}}.
\end{aligned}
\end{equation}
When $\omega_0=\Omega_0$, the interacting Hamiltonian is time-independent, which can be written as:
\begin{equation}\label{eq:}
\begin{aligned}[b]
e^{iht}H_{int}e^{-iht}&\approx\sum_{k}R\ketbra{t_{-k}}{0}+R^*\ketbra{0}{t_{-k}}\\
&=\sum_{k}Rt^{\dagger}_{-k}t^{\dagger}_k+R^*t_kt_{-k}.
\end{aligned}
\end{equation}
If we shift the triplon energy $\omega_k\to\omega_k-\Omega_0/2$, the time translation operator will be:
\begin{equation}\label{eq:}
\begin{aligned}[b]
e^{iht}=e^{i(\omega_0-\frac{\Omega_0}{2})t}\sum_k\ketbra{t_k} +e^{-i\frac{\Omega_0}{2}t}\ketbra{0},
\end{aligned}
\end{equation}
which does not affect the underlying physics. Thus, the total Hamiltonian becomes:
\begin{equation}\label{eq:}
\begin{aligned}[b]
H=\sum_{k}\tilde{\omega}_{k} t^{\dagger}_{k}t_{k}+\sum_{k}(Rt^{\dagger}_{-k}t^{\dagger}_k+R^*t_kt_{-k}).
\end{aligned}
\end{equation}
With Heisenberg equation of motion, we have the time evolution of the operator $t_k$:
\begin{equation} 
\begin{aligned}[b]
i\hbar \frac{dt_k}{dt}&=[t_k, H]=\tilde{\omega}_kt_k+2Rt^{\dagger}_{-k},\\
i\hbar \frac{dt^{\dagger}_{-k}}{dt}&=[t^{\dagger}_{-k}, H]=-\tilde{\omega}_{-k}t_{-k}-2R^*t_{k},
\end{aligned}
\end{equation}
where $\tilde{\omega}_k=\omega_k-\Omega_0/2$. We can rewrite it in a compact form:
\begin{equation}\label{eq:}
\begin{aligned}[b]
i\hbar\frac{d T_k(t)}{dt}=M_k T_k(t),
\end{aligned}
\end{equation} 
where $T_k(t)=(\langle t_k \rangle, \langle t^*_{-k} \rangle)$ is the classical amplitudes of the triplon fields, and $M_k$ is the dynamical matrix with dissipative damping, which can be written as:
\begin{equation}\label{eq:}
\begin{aligned}[b]
M_k=
\begin{pmatrix}
\tilde{\omega}_k -i\frac{\gamma}{2} & 2R \\
-2R^* & -\tilde{\omega}_{-k}-i\frac{\gamma}{2}
\end{pmatrix}.
\end{aligned}
\end{equation} 
One can obtain the eigenvalues of the dynamical matrix: 
\begin{equation}\label{eq:}
\begin{aligned}[b]
\omega_{k,\pm}=\frac{\tilde{\omega}_k-\tilde{\omega}_{-k}}{2}-\frac{i\gamma}{2}\pm\sqrt{\frac{(\tilde{\omega}_k+\tilde{\omega}_{-k})^2}{4}-4\abs{R}^2}
\end{aligned}
\end{equation} 
The $\tilde{\omega}_k+\tilde{\omega}_{-k}=\Delta_k$ is the detuning term.
\end{document}